\documentclass[]{iopart}

\usepackage{graphicx,epsfig}
\begin{document}


\def\be{\begin{equation}}
\def\ee{\end{equation}}
\def\bea{\begin{eqnarray}}
\def\eea{\end{eqnarray}}
\def\degree{$^{\circ}$}

\title[]{Inclusion of Experimental Information in First Principles Modeling of Materials}

\author{Parthapratim Biswas\footnote[1]{ 
biswas@phy.ohiou.edu}, De Nyago Tafen\footnote[2]{tafende@helios.phy.ohiou.edu}, 
Raymond Atta-Fynn\footnote[3]{attafynn@helios.phy.ohiou.edu}, and David Drabold
\footnote[4]{drabold@ohio.edu}}

\address{Department of Physics and Astronomy, Ohio University, Athens, OH 45701, USA} 

\begin{abstract} 
We propose a novel approach to model amorphous materials using a first principles 
density functional method while simultaneously enforcing agreement with selected 
experimental data. We illustrate our method with applications to amorphous silicon 
and glassy GeSe$_2$. The structural, vibrational and electronic properties of the 
models are found to be in agreement with experimental results. The method is general 
and can be extended to other complex materials.  
\end{abstract}




\section{Introduction}

Michael Thorpe has devoted an important part of his distinguished 
career to exploring the structure of materials.  For this reason, 
it is not inappropriate for us to contribute a novel technique for 
forming models of complex materials to this volume. We offer this 
work with respect and affection to Mike, in honor of his 60th birthday. 
An essential feature of materials science is the synergy between 
experiment and theory. A properly conducted experiment reveals the 
fundamental characteristics of a material, be they structural, 
electronic, magnetic or other. For cases in which there is no doubt 
of the reliability of an experiment, the data hold primacy over any 
theoretical considerations -- the data express what we know to be 
true about the material.  Experiments, however, provide {\it highly 
incomplete} information about the system in question. Virtually every 
experimental technique probes large, and often macroscopic volumes of 
a sample. On the other hand, many properties (particularly electronic 
and optical properties) may be determined by tiny fractions of the 
sample (those parts associated with defects, for example), which are 
often invisible to the probe, or if visible, averaged over all the 
defects in the material. Simulation has the great virtue that it provides 
exquisite atomistic detail -- atomic coordinates, localized electronic 
eigenstates, vibrational eigenmodes {\it etc.}, but  any simulation 
scheme requires {\it many} approximations (typically including unphysically 
short time scales, small models and always approximate interatomic 
interactions). In practice, the modeler requires experimental information 
to discern whether his or her model is credible.  The current {\it modus 
operandi} is then to adopt a simulation scheme, make a model and hope 
for the best.  If there are no serious contradictions with experiment 
the work is published and new insights may accrue. Many successes have 
come from this approach to modeling, but it is often the case that 
contradictions with experiment do exist, in which case the model is not 
precisely describing nature.  In this paper, we make a first attempt 
to {\it systematize} the process of forcing atomistic models to be maximally 
consistent with experimental data {\it and} appropriate interatomic 
interactions, and offer a recipe to accomplish this.  With the importance 
of experimental information clearly in mind, an alternative paradigm has 
been proposed by McGreevy and others, the so-called ``Reverse Monte 
Carlo" (RMC) method~\cite{rmc4,gereben,walt}. Here, one sets out to 
build an atomistic model agreeing with experiment(s).  This is implemented 
by making Monte Carlo moves of atomic coordinates which drive a structural 
model toward exact agreement with one or more experiments. No interatomic 
interaction is needed for the method. The logic of the method is simple 
and appealing. 

In this paper, we merge {\it {ab initio}} molecular relaxation (MR) with  
RMC. What makes this hybrid approach attractive is that the methods (MR 
and RMC) are complementary to each other.  One can describe this scheme 
as a way to ``tune" a structural model using MD within the space of 
{\it {experimentally realistic}} models as defined by RMC. In this paper, 
we choose two materials to illustrate our scheme: the first one is amorphous 
silicon which is notoriously hard to model by melt quenching, whereas the 
second one is a troublesome and complex material with a long experimental 
and modeling history: g-GeSe$_2$.

\section{Summary of Reverse Monte Carlo Simulation}

RMC has been described in detail elsewhere~\cite{rmc4}. At its simplest, it 
is a technique for generating structural configurations based on experimental data. 
The method was originally developed by McGreevy \& Pustazi~\cite{rmc4,gereben} 
for liquid and glassy materials for lack of different routes to explore 
experimental data but in recent years progress has been made toward modeling 
crystalline systems as well. Starting with a suitable configuration, 
atoms are displaced randomly using periodic boundary conditions until the 
input experimental data (either the structure factor or the radial distribution 
function) match the data obtained from the generated configuration. This is 
achieved by minimizing a cost function which consists of either structure 
factor or radial distribution function along with some appropriately chosen
constraints to restrict the search space. Consider a system having $N$ number 
of atoms with periodic boundary condition. One can construct a generalized 
cost function for an arbitrary configuration by writing :
\be
\label{eq1}
\xi=\sum_{j=1}^{K} \sum_{i=1}^{M_K} \eta_{i}^j \{F^j_E(Q_i)- F^j_c(Q_i)\}^2
   + \sum_{l=1}^L \lambda_l P_l
\ee
\noindent

where $\eta_{i}^j$ is related to the uncertainty associated with the 
determination of experimental data points as well as the relative weight 
factor for each set of different experimental data. Here $K$ and $M_K$ 
stands for the total number of different experimental data set and 
data points.  The quantity $Q$ is the appropriate generalized variable 
associated with experimental data $F(Q)$ and $P_l >0$ is the penalty 
function associated with each constraint (if minimized to zero implies
exact agreement with experimental data set $l$), and $\lambda_l$ is a 
(positive) weight factor for each constraint.

\section{Experimentally Constrained Molecular Relaxation }
\begin{figure}
\includegraphics[width=2.8 in, height=3.0 in, angle=0]{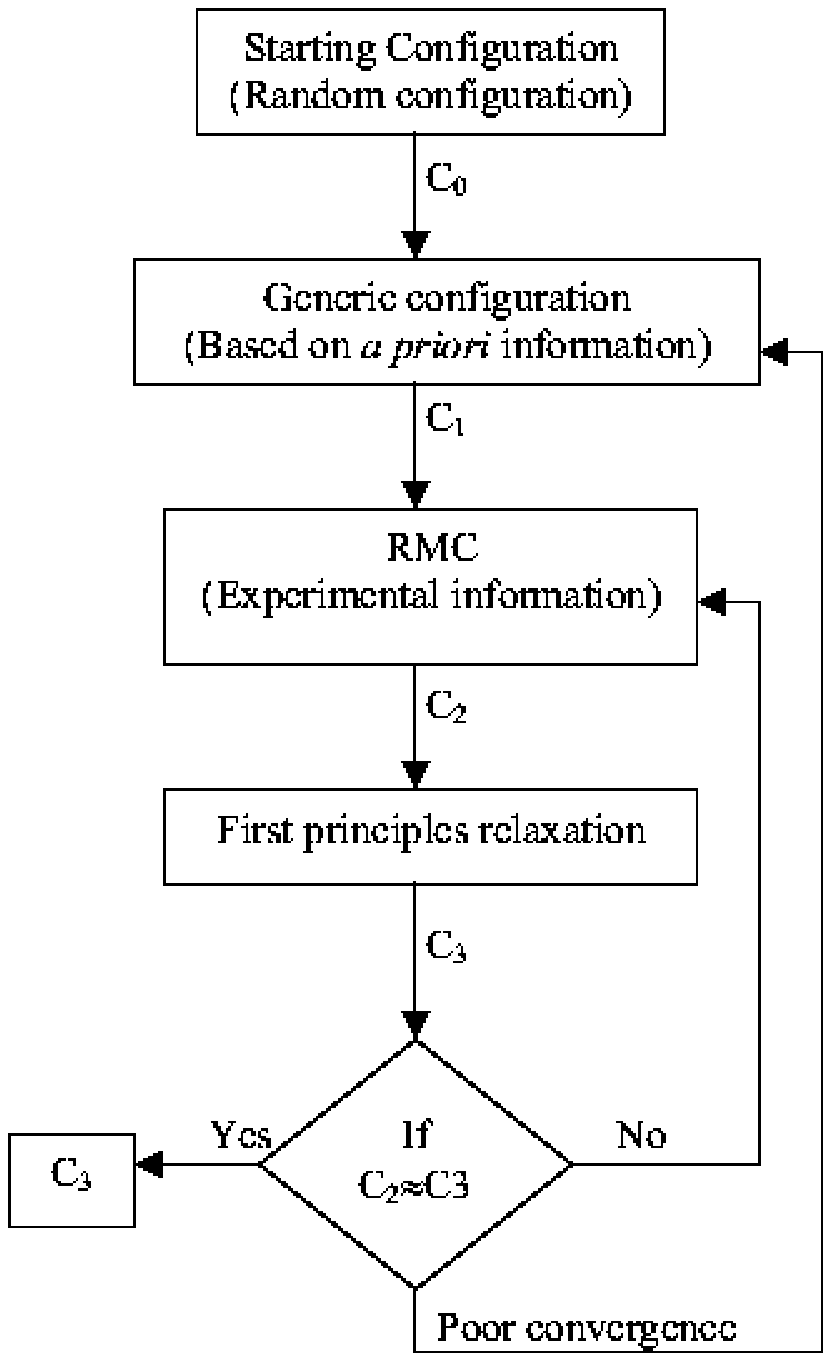}
\includegraphics[width=2.6 in, height=2.8 in, angle=0]{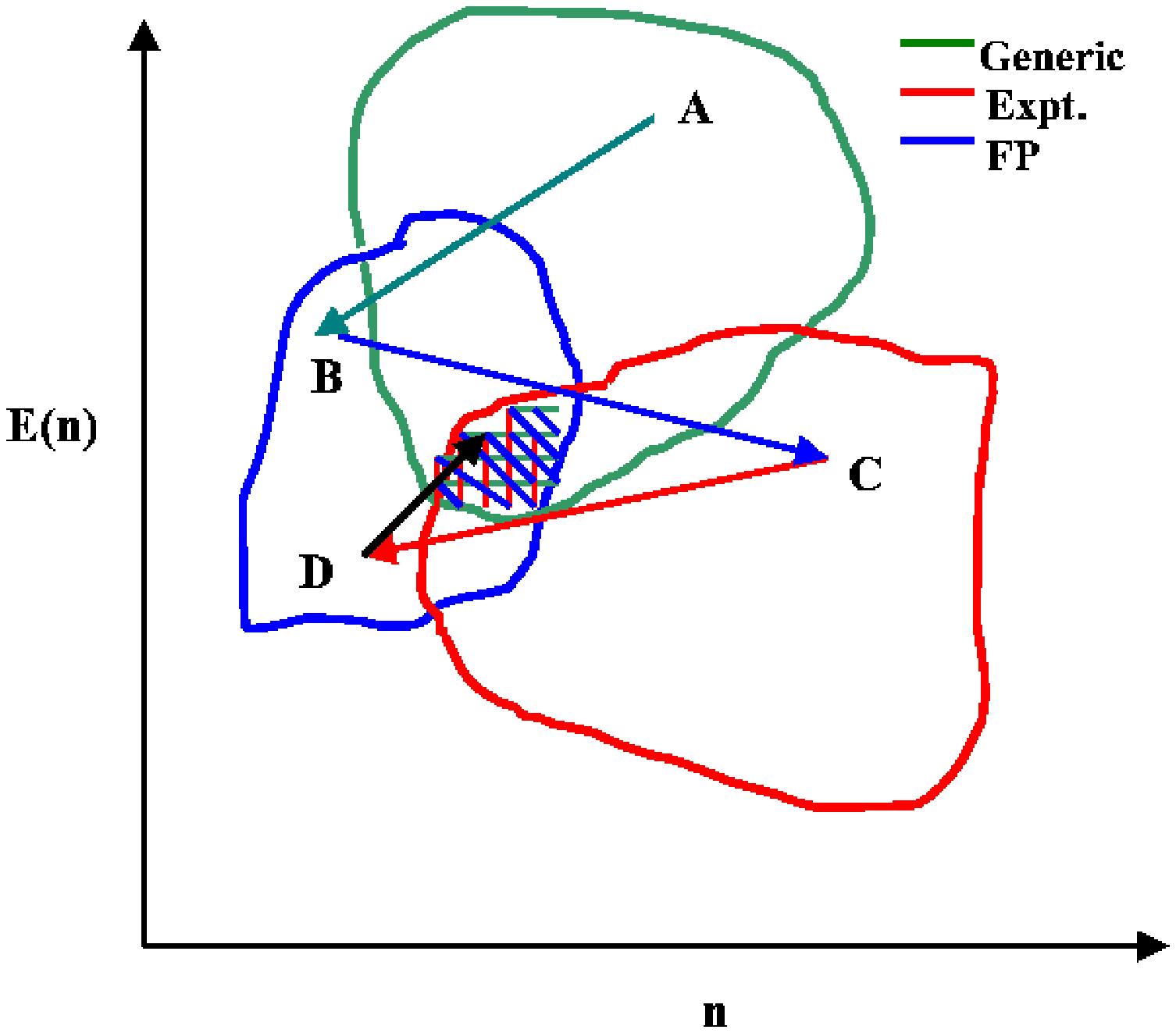}
\caption{ 
(Left panel) A flow diagram showing the different steps used in our 
ECMR method.  \\
(Right panel; Color online) A cartoon diagram of configuration space showing how the method 
works. Those parts of the configuration space consistent with experimental and first principles (FP) energy minima 
are indicated along with a ``generic" space, which includes only some 
primitive {\it a priori} constraints. The cross-hatched region is 
consistent with both experiments and theory which corresponds to realistic 
configurations of materials. ECMR can drive a starting configurations to 
this region following a possible path shown in the figure.
} 
\label{fig1}
\end{figure}
From the preceding section, it is apparent that inclusion of adequate 
information is crucial for successful production of a realistic (representative) configuration. 
Ideally, one should enforce {\it all} the experiments available.
Typically this might include structural, electronic and vibrational
properties for a well-studied material. One possible way to keep the problem tractable is to use 
minimal experimental information in conjunction with 
a knowledge of interactions between the atoms. The obvious means to do this 
is to add a constraint in RMC (an additional $P_l$ in Eq.~\ref{eq1}) to minimize the magnitude of the force 
on all the atoms according to some energy functional or possibly to 
minimize the total energy. Particularly for an {\it ab initio} Hamiltonian 
this is expensive, since Monte Carlo minimization of Eq.~\ref{eq1} requires 
a large number of energy/force calls.  Thus, we instead employ a simple 
``self consistent" iteration scheme: (1) starting with an initial 
``generic" configuration~\cite{note2} C$_1$, minimize $\xi$ 
to get C$_2$, 2) Steepest-descent 
quench C$_2$ with an {\it ab initio} method to get C$_3$, (3) subject the 
resulting configuration to another RMC run (minimize $\xi$ again), repeat 
steps (2) and (3) until both the force-field relaxed model and RMC models no longer 
change with further iteration. For the RMC component of the iteration, we 
make the conventional choice of using Monte Carlo for the minimization. 
This is easily implemented and does not require gradients (and thus allows the use of 
non-analytic terms in Eq.~\ref{eq1}~\cite{gradients}, if desired).  A schematic 
representation of our method is presented in Figure~\ref{fig1}, and a cartoon
suggesting a possible trajectory through configuration space (starting from the
broadest ``generic network" space, and ideally concluding in the cross-hatched
region representing a minimum both for experiment and for accurate interatomic
interactions). Since the 
method directly uses experimental information in association with first principles 
relaxation, we refer to this method as experimentally constrained molecular 
relaxation or ECMR. 

We emphasize that our method is {\it flexible}. The logic of our method
suggests that one should include whatever important experimental information is 
available. In this paper we limit ourselves to the pair-correlation functions. 
In principle, other experiments could be included as well. These might be 
costly to include (for example to compel agreement with the vibrational 
density of states, the dynamical matrix would be required at each iteration). 
The method is equally suited to fast empirical potentials, which would allow 
studies of very large models. It is also easy to force a close fit to some 
restricted range of data, and a less precise fit elsewhere if desired. Our 
scheme also provides insight into the topological signatures of different 
constraints (experimental or otherwise). Chemical and or topological constraints 
could also be maintained as part of the RMC iteration.

Our method can  be understood as a way to minimize an effective potential energy 
function $V_{\mbox{eff}\,}(R) = V(R) + \Lambda \zeta(R)$, where $V(R)$ is the 
potential energy of the configuration (denoted by $R$), $\Lambda >0$, and 
$\zeta$ is a non-negative 
cost function enforcing experimental (or other) constraints as in Eq.~\ref{eq1}. 
Empirically, we find that it is possible to find configurations that simultaneously 
approximately minimize both terms (which implies that the choice of $\Lambda$ is 
not very important). It is also clear that our method is really {\it statistical}: in 
general one should generate an ensemble of conformations using ECMR. 

\section{ECMR model of glassy GeSe$_2$}

In this section, we apply our ECMR method to g-GeSe$_2$, a 
classic glass former. The material is particularly known for its interesting 
physical properties: (1) it displays nanoscale order: a ``first sharp diffraction 
peak" (FSDP) is observed in neutron diffraction measurement, (2) the material has 
interesting photoresponse (understanding of which requires the electronic 
structure), (3) the material is difficult to simulate with {\it ab initio} 
techniques~\cite{Massobrio 1998, Tafen 2003,Cobb 1996,Zhang 2000}. 

\begin{figure}
\includegraphics[width=3.0 in, height=2.8 in, angle=270]{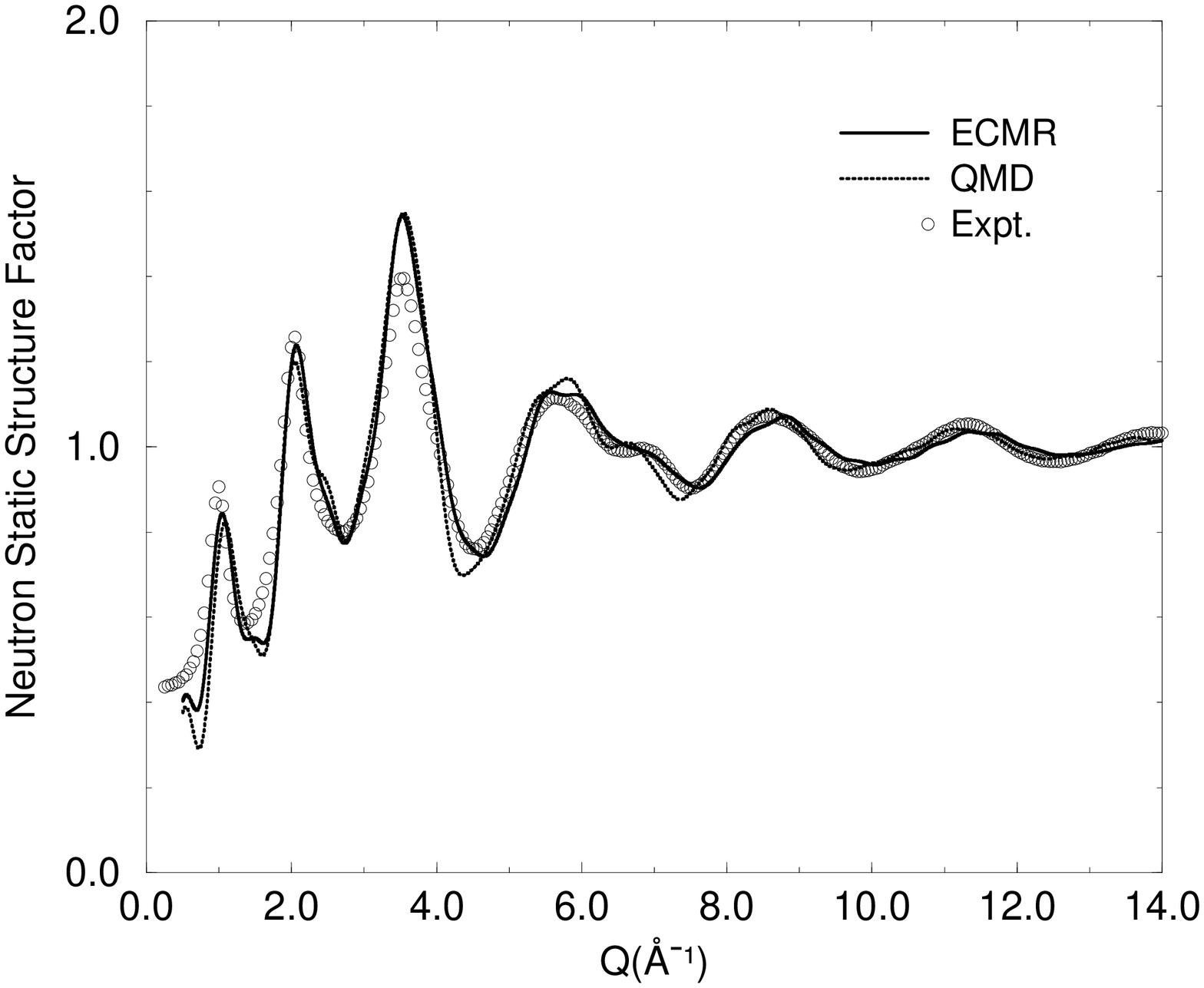}
\includegraphics[width=3.0 in, height=2.8 in, angle=270]{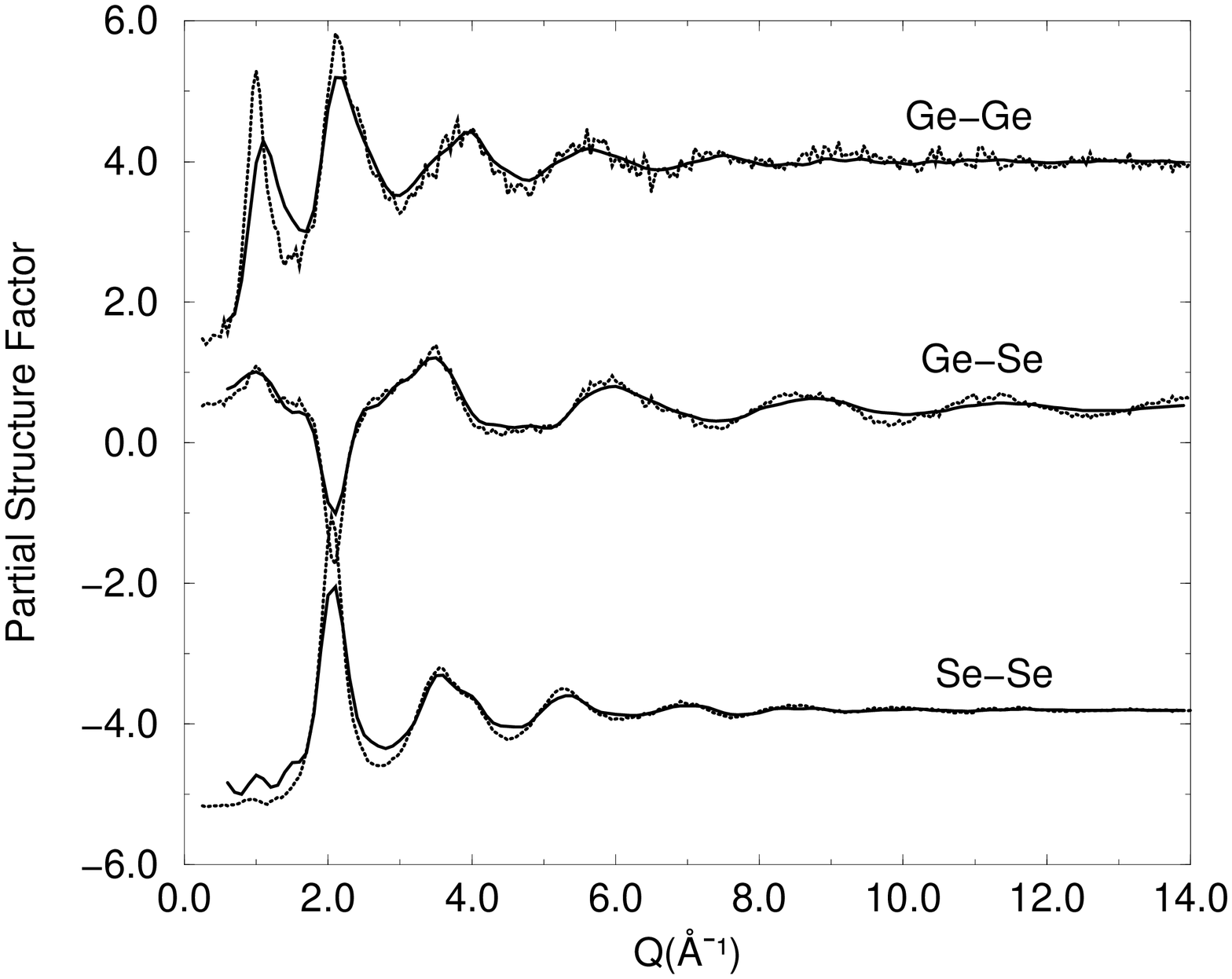}
\caption{
\label{fig2}
(Left panel) The total static structure factor obtained from Neutron diffraction 
experiments\cite{Petri 2000}  (open circle) and the present work (solid [converged ECMR] and dotted 
[converged MD] lines). The first sharp diffraction peak (FSDP) agrees closely with 
the result obtained from our ECMR model. \\
(Right panel) The partial structure factors of the ECMR model along 
with the experimental data\cite{Petri 2000}  are plotted for a direct comparison. 
}
\end{figure}

\begin{figure}
\includegraphics[width=2.5 in, height=2.2 in, angle=270]{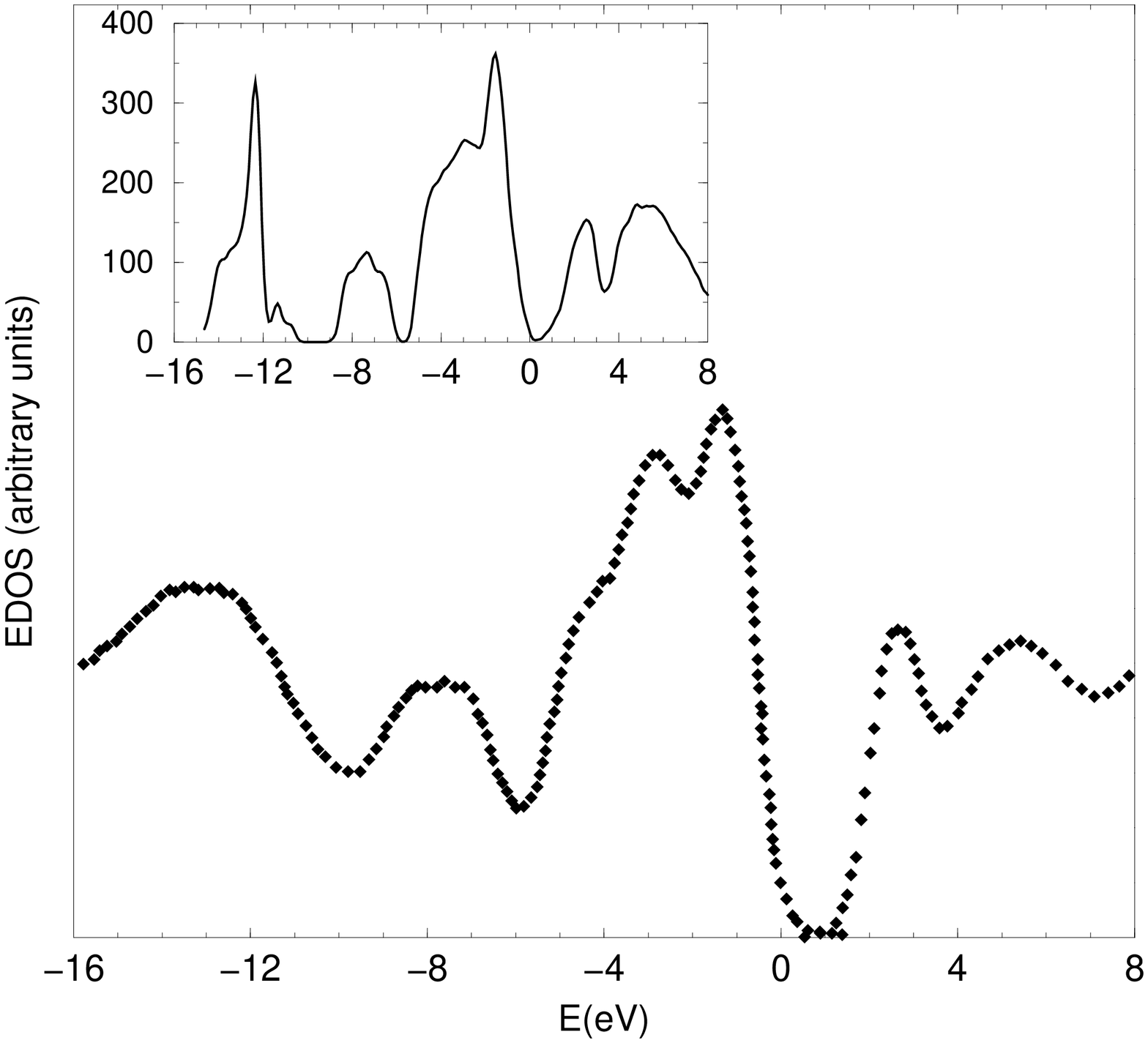}
\includegraphics[width=2.7in, height=2.8in,angle=270]{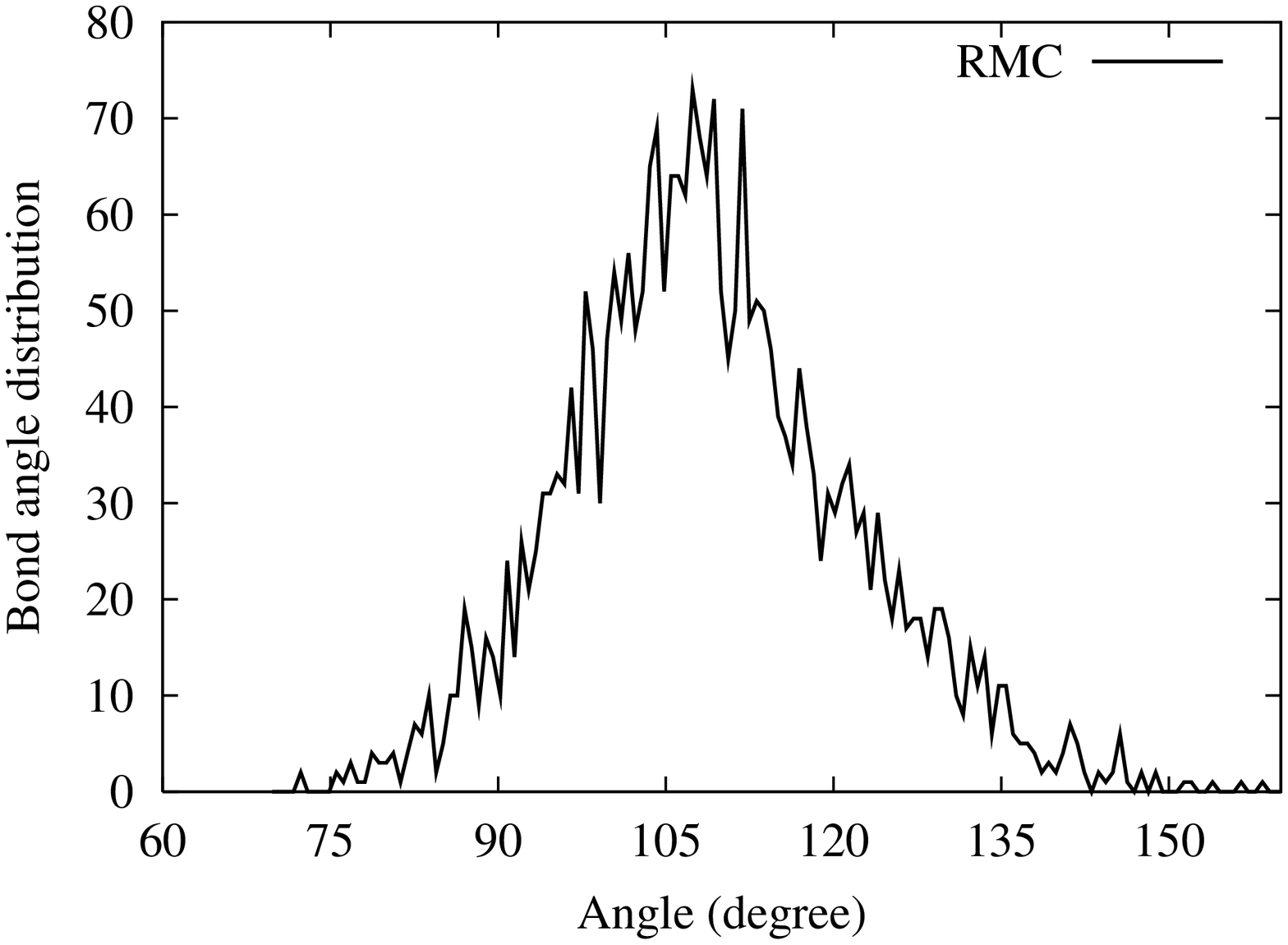}
\caption{
(Left panel) The electronic density of states (Gaussian-broadened Kohn-Sham
eigenvalues) for ECMR model of GeSe$_2$(inset). The XPS~\cite{Bergignat 1988} and
IPES~\cite{Hosokawa 1994} data show the occupied (valence band) and unoccupied 
(conduction band) part of the spectrum. The Fermi level is at E=0.\\
(Right panel) The bond angle distribution for the RMC model of 
amorphous silicon described in the text. The average bond angle 
is found to be 109.01{\degree} with root-mean-square deviation 
12.5{\degree}. 
}
\label{ge-edos}
\end{figure}

The model used in our calculation consists of 647 atoms of Ge and Se in a cubic 
box of size 27.525 {\AA}. We start with a configuration (C$_1$ in figure\ref{fig1}) 
obtained by constraining the coordination number (2 for Se, 4 for Ge) and 
the bond-angle distribution of Se-Ge-Se to an approximate Gaussian with the 
average bond angle 109.5{\degree}. The configuration is ``generic" in the sense 
that it includes none of the detailed chemistry of Ge and Se aside from the 
8-N rule and initially heteropolar bonding. Salmon and Petri~\cite{Petri 2000} have recently measured separately the three 
(Ge-Se, Ge-Ge and Se-Se) partial structure factors using the method of 
isotopic substitution. We utilized all  three partials (in real space) 
in the RMC step to construct an approximate configuration. The MD relaxation was 
done with FIREBALL~\cite{fireball}. It was found that after the fourth ECMR iteration, 
S(Q) hardly changed. In figure~\ref{fig2}, we plot the experimental partial
structure factors, and the total S(Q) after the fourth iteration of ECMR relaxation. 
The agreement between the ECMR obtained partials structure factors and the experiment 
is very reasonable. The difference is possibly due to relatively small system size and 
small inconsistencies between the experiments and the total energy functional.  The FSDP 
in figure~\ref{fig2} is genuinely well reproduced (very close in width and centering, 
and much improved from all previous models in height). Moreover, as 
for our ``Decorate and Relax" method~\cite{tafen}, the large $Q$ structure closely 
tracks experiment (unlike the experience for quench from the melt models). These 
features are of course ``built in"; we show here that the method does preserve every 
important feature of the structure of the glass manifested in S(Q). 
A characteristic feature of GeSe$_2$ glass is the presence of edge- and 
corner-sharing tetrahedra. Raman spectroscopy~\cite{Jackson 1999} and 
Neutron diffraction~\cite{Susman 1990} studies have indicated that almost 
33\% to 40\% of total Ge atoms are involved in edge-sharing tetrahedra. 
The fraction obtained from our model is found to be 38\%. We also 
have observed that about 81\% of Ge atoms of our model are 4-fold
coordinated of which approximately 75\%  form predominant Ge-centered
structural motifs $Ge(Se_{\frac{1}{2}})_4$ while the rest 6\% are
ethane-like $Ge_2(Se_{\frac{1}{2}})_6$ units.  The remaining Ge
atoms are 3-fold coordinated and are mostly found to be bonded
as Ge--Se$_3$ units. On the other hand, the percentage of 2-, 3-
and 1-fold  coordinated Se atoms are 72\%, 18\% and 10\% respectively.
M\"ossbauer experiments, where Sn was used as a Ge
probe~\cite{Boolchand 1982}, estimated that the fraction of Ge
involved in dimers is 16\%. Petri and Salmon~\cite{Petri 2000} have 
recently studied local atomic environment by Neutron diffraction 
measurements. By integrating partial radial distribution functions 
via Fourier transform of structure factors
they obtained nearest neighbor coordination numbers  $n_{Ge-Ge}$ = 0.25,
$n_{Se-Se}$ = 0.20, and $n_{Ge-Se}$ = 3.7 that corresponds to average
coordination number $\bar n $ = 2.68. The corresponding values from
our model are : $n_{Ge-Ge}$ = 0.17, $n_{Se-Se}$ = 0.30, $n_{Ge-Se}$ = 3.68,
and $\bar n $ = 2.66. The partial and total coordination numbers, therefore,
agree well with experiments and are consistent with the 8-N rule which
predicts $\bar n$ = 2.67. The percentage of homopolar bonds present in our
model is found to be about 6.2 \% which is again very close to the value
8\% noted by Petri and Salmon~\cite{Petri 2000}.
In figure \ref{ge-edos}, we have plotted the electronic density of states (EDOS) 
of our ECMR model using a first principles density functional code 
FIREBALL~\cite{fireball}. The EDOS is obtained by convoluting 
each energy eigenvalue with suitably broadened Gaussian.  Our EDOS in figure~\ref{ge-edos} 
agrees well with experimental results obtained from x-ray photo-emission
spectroscopy~\cite{Bergignat 1988} (XPS), inverse photo-emission 
spectroscopy\cite{Hosokawa 1994} (IPES) and ultraviolet photo-emission
spectroscopy~\cite{Hino 1980} (UPS) measurements as well
as with those obtained in recent theoretical studies~\cite{Louie 1982, Pollard 1992,
Cobb 1996, Tafen 2003}. Interestingly, all the relevant experimental features are
also found in the calculated EDOS, providing further support for the reliability
and accuracy of our method. The substantial splitting between the first two
peaks of the valence bands named the \textit{$A_1$} and \textit{$A_2$} peaks
is well pronounced. 

Finally, we examine the vibrational density of states (VDOS) of our 
model due to the close relationship to its atomic-scale structure 
and its mechanical properties. The computational methodology was 
reported elsewhere~\cite{Cobb 1996}. Comparing our VDOS with experiment 
obtained by inelastic neutron-scattering~\cite{Cappelletti 1995}, the 
spectrum exhibits the same features. Three bands can be distinguished: 
a low energy acoustic band involving mainly extended interblock 
vibrations and a high energy optic band consisting of more localized 
intrablock vibrations. The two main bands are found to be clearly 
separated by the tetrahedral breathing (\textit{A$_1$}-\textit{A$_{1c}$}) 
band. The overall agreement is rather good. The 2-4 meV redshift from experiment starting
at 35 meV and the discrepancy around 17-20 meV have also been
observed by Cobb et al~\cite{Cobb 1996}. The \textit{A$_1$--A$_{1c}$} 
splitting is difficult to determine accurately from the vibrational 
density of states. In this work, we have observed two peaks in the 
\textit{A$_1$--A$_{1c}$} region which are separated by approximately 
2.5meV to 3.5meV.

\section{RMC model of Amorphous Silicon}

Having studied g-GeSe$_2$ we now address amorphous silicon, an important 
material with a wide variety of applications. The structure of a-Si is 
well represented by continuous random network (CRN) model.  A number of 
techniques have been developed over the last three decades to produce 
CRN networks, but the jewel of all these is undoubtedly the so-called 
``bond switching" algorithm of Wooten, Winer and Weaire (WWW)~\cite{woot,ww}. 
Other methods such as quench from the melt simulation, while working for a number 
of glasses (e.g. a-SiO$_2$), is found to be far less effective for modeling 
a-Si.  The ``quench from the melt" produces too much liquid 
character and high defect concentration that lead to incorrect topology of 
the network. Although there is no definitive reason as to why it is so 
difficult to model from the melt, the failure is generally attributed to 
the presence of 6-fold coordinated atoms in liquid Si in contrast to 
amorphous state where tetrahedral bonding is considered as one of the 
characteristic features of the network. One supposes that MD within its 
short time scale cannot explore the entire phase space to determine the 
tetrahedral network topology of the amorphous silicon. Reverse Monte Carlo 
has been applied to model a number of materials including amorphous 
silicon~\cite{rmc4}. Of particular interest related to a-Si 
are the models developed by Gereban and Pusztai~\cite{gereben} and Walters and 
Newport~\cite{walt}. Although these early models are not useful in 
studying electronic properties, they do contain some characteristic 
features of tetrahedral bonding of a-Si network.

The particular RMC scheme that we implement can be found in Reference~\cite{biswas}. 
We compel agreement with the static structure factor to keep the bond angles 
as close as possible to the tetrahedral value, while also maintaining four-fold 
coordination. Since we are interested in both the structural and electronic properties, 
we choose to model a reasonable system size of 500 atoms in a cubic box of length 
21.18{\AA}. The results for the RMC model (obtained after RMC fit 
followed by first principles relaxation using {\sc Siesta} 
Harris~\cite{siesta}) are presented in figure~\ref{ge-edos}--\ref{mro}.
The structure factor for the final configuration is plotted in the 
left panel of figure~\ref{mro} along with the experimental data 
obtained by Laaziri et al~\cite{laaziri}. It is clear from the 
plot that the experimental data agree closely with our RMC model  
except for the few points near the first peak. It is tempting to suggest 
that this (small) deviation is possibly due small size of our model. 
But, we have noticed that this deviation continues to persist even in large, 
high quality continuous random network generated by WWW algorithm. 
Next we focus on the bond angle distribution of our model.  Structure 
factor or radial distribution function can only provide information at 
the two-body level. A further characterization of the network is 
therefore necessary in terms of higher order correlation functions which 
requires examining at least bond angle distribution of the network. 
The average bond angle and the root-mean-square (RMS) deviation provide 
a good measure of the quality of the CRN network which is plotted 
in the right panel of figure~\ref{ge-edos}. The average bond angle of 
our RMC model is found to be 109.01{\degree} along with the RMS 12.5{\degree}. 
The latter is slightly higher than its WWW counterpart where an RMS deviation 
as narrow as 9.9{\degree} has been reported~\cite{bark}. 
The number of four-fold coordinated atoms is 88\% while 
the remaining 12\% consists of three- and five-fold corodinated atoms. 
We also observe that the characteristic features of a-Si electronic DOS 
are correctly produced although the bandtailing is exaggerated. We are 
working on implementing ECMR for a-Si which we believe will address 
some of the existing problems with our present RMC model.

\section{ECMR and Medium range Order}

Despite the lack of long-range translational and orientational order, 
some
covalent amorphous solids exhibit structural order (with varying degree)
at the medium length scale. MRO is associated with length scales of few 
nanometers, and may affect mechanical, electronic and optical properties 
of materials.  Diffraction measurements
provide little information at this length scale because of the
isotropic nature of most disordered materials.

Recent development of Fluctuation electron microscopy (FEM) provides an
opportunity to detect medium-range order in non-crystalline materials~\cite{treacy}.
FEM is essentially an electron microscopy technique that can
be used to detect MRO by measuring fluctuations in the diffracted 
intensity
originating from the nano-scale volumes in the sample. Thus, the FEM
signal carries information about  the MRO and
by gathering information from these different nano-scale volumes MRO 
can be
detected by computing the normalized variance of the diffracted beam 
intensity.
The meaning of the FEM signal has been recently 
developed by
Treacy and Gibson~\cite{treacy} which shows that the FEM signal is 
sensitive to two-, three- and four-body correlation functions of 
the sample and contain information
which is absent in a standard diffraction measurement. The normalized 
variance of the measured intensity can be written as~\cite{voyles1},
\[
V(k) = \frac{\langle I^2(k, Q)\rangle}{{\langle I(k, Q) \rangle}^2} - 1 
\] 
where $\langle I \rangle$ is the measured intensity, $k$ is the scattering 
vector, $Q$ is inversely proportional to the real space resolution which is 
usually set to the length scale at which MRO is believed to exist and 
$\langle \rangle$ indicates averaging over the positions of the sample. 
Recent studies of amorphous thin films via FEM have led to the suggestion 
that such films of a-Si and a-Ge contain small highly strained ordered 
crystalline grains known as paracrystals. These paracrystalline and continuous 
random network models can have structure factor that are almost indistinguishable 
with difference occurring only at medium-range length scales.  Computer 
modeling has recently indicated that certain non-crystalline Si films may not 
be simply a continuous random network, but rather consist of nano-sized strained
crystalline grains embedded in a continuous random network matrix~\cite{voyles}.
While  such nano-sized crystalline grains in a CRN matrix is sufficient to 
explain the FEM data, some questions remain unanswered. The question of 
whether the paracrystalline model is the only one containing the signature of 
MRO is still a matter of conjecture. Since we know from RMC simulation that one 
can generate configurations of a-Si having almost identical structure factor
but with very different topology, it is natural to explore the possibility of 
constructing CRN models that do not  contain nano-sized grains in the model and 
yet displays the characteristic signature of MRO via FEM data. Motivated by 
the success of our ECMR approach to model glassy GeSe$_2$, we recently attempted 
to include MRO in a CRN model via following scheme: (1) construct a cost 
function comprises of FEM signal and radial distribution function (2) minimize 
the cost function via RMC subject to the constraint that no atoms can come closer
to 2.0{\AA} and, (3) accept only those moves during RMC simulation (in step 2)
that only lower the total energy of the configuration in order to
generate a stable, minimally strained network. 

The cost function is similar to the one we have used earlier except that
we include experimentally obtained FEM data along with the radial 
distribution function. We start with a cubic supercell (box length 
43.42{\AA}) consisting of 4096-atom model of a-Si due to Djordjevic, 
Thorpe and Wooten~\cite{dtw}. This initial configuration does not 
have any signature of MRO as is indicated by a flat FEM signal. Using 
RMC scheme, we match experimental FEM data, permitting only
moves that simultaneously minimize both the cost function and the
total energy of the system obtained from a modified Stillinger-Weber 
potential~\cite{msw}. The final ECMR relaxed configuration is 
found to be 100\% four-fold coordinated and has energy per atom 
-3.007 eV. This is significant considering the fact that the lowest 
energy configuaration obtained after relaxing the initial configuration 
(again with the modified  Stillinger-Weber potential) has energy -3.013 eV/atom. 
The structure factor and the electronic density of states of the 
fitted CRN are found to agree quite well with experiments.  

\begin{figure}
\includegraphics[width=2.7in, height=2.6in,angle=270]{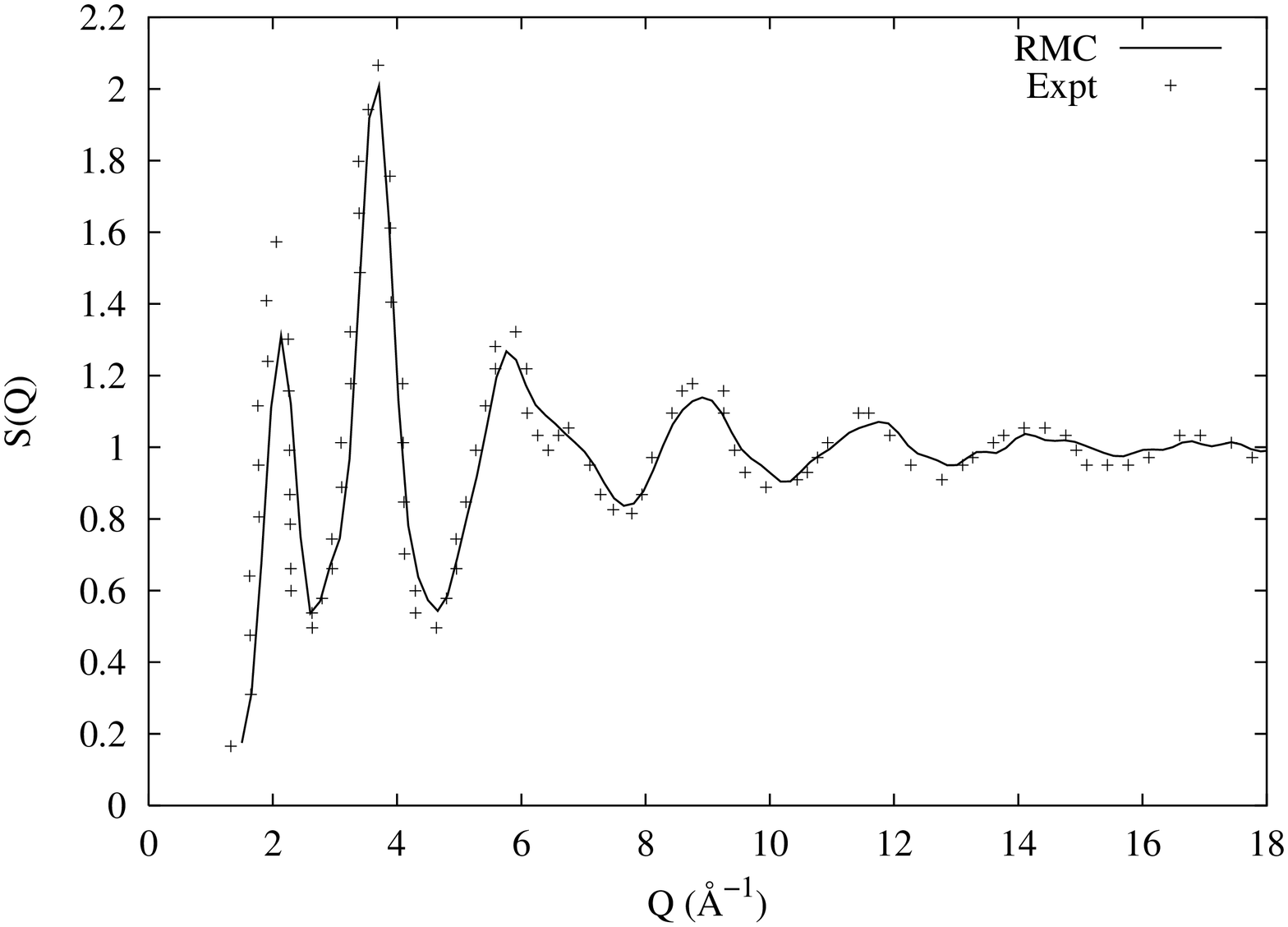}
\includegraphics[width=3.4in, height=3.0in,angle=270]{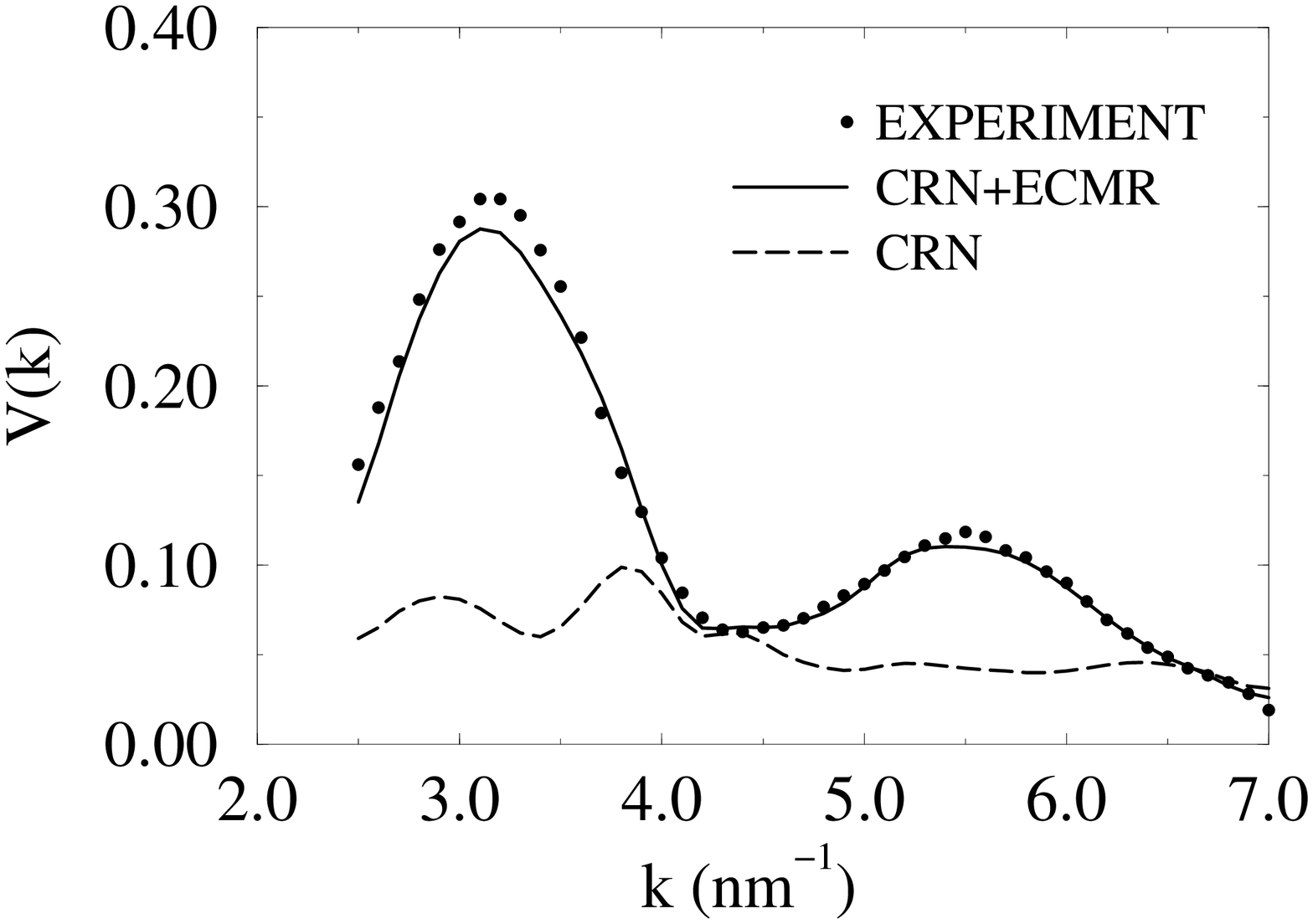}
\caption{\label{mro} 
(Left panel) Static structure factor for the RMC model of amorphous silicon 
along with the experimental data of Laaziri et al~\cite{laaziri}. \\
(Right panel) Image intensity variance ($V$) as a  function of scattering 
vector magnitude ($k$) for a 4096-atom ECMR-CRN and a CRN model along with 
the experimental data of as-deposited {\it a}-Si measured by Voyles et 
al~\cite{voyles}. The resolution for the experimental and simulated 
(both starting and final) data are fixed at 15 {\AA} and 10 {\AA} 
respectively. The experimental FEM data are multiplied by 100 before the 
start of simulation. 
}
\end{figure}

In FEM, one estimates medium-range order by studying images obtained 
(from FEM) in two different modes: variable resolution (VRFEM) and 
variable coherence (VCFEM). In variable resolution one measures the 
characteristic MRO length scale by varying $Q$ for a fixed value 
of $k$, while variation of $k$ for a given $Q$ yields the information 
about medium-range structure. In figure~\ref{mro}, we have plotted 
the variable coherence FEM data for both the starting and final 
continuous random network configurations. For the purpose of comparison, 
we have also plotted the experimental data obtained by Voyles et 
al~\cite{voyles}. We choose a low resolution to probe atomic correlation 
within a volume of 10 {\AA} radius. The presence of MRO is manifested 
in the form a large variance and strong $k$ dependence of the VCFEM signal.  
Figure~\ref{mro} clearly reveals this feature and shows that the 
fit with the experimental VCFEM data is very good. This shows that 
our simulated CRN model not only has correct short-range structure 
(reflected in radial and bond angle distributions) and electronic 
density of states but also displays a strong FEM signal which confirms 
the presence of medium-range order in the sample.  

An intriguing feature of MRO (appearing in VCFEM signal) is the 
position of the peaks at $k$ values 0.3 {\AA} and 0.55 {\AA}. The 
atomistic origin (in terms of real space MRO structure) of these peaks 
is still not very clear. In the paracrystalline model, 
these maxima are thought to be associated with Bragg's diffraction from 
crystalline grains and their relative orientation 
with each other. While one cannot rule out this possibility, 
there are indications that the peaks are independent of the 
presence of crystalline grains in the structure. Our preliminary 
studies of the 4000-atom paracrystalline model have shown that the peaks 
are present even when grains are removed from the model. It is therefore 
not unreasonable to suggest that the paracrystalline structure 
(i.e. the presence of small crystalline grains in a CRN matrix) while 
sufficient, is not necessary to account for the experimental data 
obtained in fluctuation microscopy. 

Finally we comment briefly on the dihedral angle distribution (DAD) 
and the presence of MRO. It is believed that dihedral angles (related 
to the reduced four-body correlation function) are another signature 
of medium-range structure in a non-crystalline network that should 
be reflected in the dihedral angle distribution. It is therefore 
pertinent to compute the dihedral angle distributions from the 
paracrystalline model, our ECMR fitted CRN, the initial CRN and 
to examine to what extent they differ from each other. Once again 
our preliminary studies have indicated that both ECMR fitted CRN 
and the initial CRN have almost identical dihedral distributions. 
As for the paracrystalline model, we have noticed a slight increase 
in the distribution function (almost negligible) near 60{\degree}
which was also reported by Voyles et al~\cite{voyles}. This result 
however must be interpreted carefully taking into consideration 
that the dihedral angle distribution in crystalline silicon is a 
sharp peak at 60{\degree}. The presence of crystalline grains 
therefore evidently favor having more dihedral angles centered 
around 60{\degree} giving an added contribution to the total 
dihedral angle distribution. 

\ack
We acknowledge the support of US National Science Foundation under 
grants DMR 0205858 and 0310933. We thank John Abelson (UIUC) for 
motivating us in this work.  \\ \\


\begin{thebibliography}{99}
\bibitem{note2} 
A ``generic" configuration does not contain any experimental information 
but empirical characteristic of the structure only (when available). For 
example, a configuration with 4-fold coordination can be treated as a generic 
configuration for $a$-Si while for g-GeSe$_2$, the coordination for Ge and 
Se should be 2 and 4 respectively. It is generic in the sense that other 
information (such as bond angle) can be used to construct such a 
configuration. One generic configuration may be favorable than another 
depending on the structure of configuration space.
\bibitem{gradients} 
It is possible that suitable gradient-based methods could 
provide more rapid convergence, a point we do not investigate 
here.
\end{thebibliography}
\end{document}